\documentstyle[12pt,times,psfig]{article}
\def\be{\begin{equation}}
\def\ee{\end{equation}}

\begin{document}
\begin{center}
{\Large\bf Random Matrix Theory with Non-integer $\beta$}\\
~\\
Peter Marko\v s\\
Institute of Physics, Slovak Academy of Sciences,
D\'ubravsk\'a cesta 9, \\
842 28 Bratislava, Slovakia\\
e-mail: fyzimark@savba.sk

\end{center}

\bigskip

\begin{abstract}
{We show that the random matrix theory with non-integer
"symmetry parameter" $\beta$  describes the statistics of transport
parameters of strongly disordered two dimensional  systems.}
\end{abstract}

\bigskip

\bigskip

The application of the random matrix theory (RMT) 
\cite{Mehta}
to electronic transport in {\sl weakly disordered} systems 
\cite{pichard}
enables us to understand the main features of the transport. 
The  small number of parameters which enters RMT explains  the universal
features of transport, especially 
the universal conductance fluctuations as well as the simple (linear) form of the spectra of
eigenvalues of the transfer matrix
\cite{imry}.

There are  only three  parameters which characterize the system in the RMT: 
the  system size $L$, 
the mean free path $l$.
and the  symmetry parameter $\beta$. 
It is believed that only three integer values
of $\beta$ have physical meaning: $\beta=1,2,4$ for orthogonal, unitary and
symplectic symmetry, respectively. Recently, however, Muttalib and Klauder
\cite{MutK} showed that the non-integer values of $\beta$ are also consistent
with the DMPK equation 
\cite{DMPK}. 
They introduced  a new parameter $\gamma$,
which substitute $\beta$  in the DMPK equation.  
$\gamma$ is determined by
statistical properties of eigenvectors and eigenvalues of the transfer matrix. 
Consequently,  it could possess any positive real value, although 
the physical symmetry of the system remains unchanged.
This observation indicates that the electronic transport in  disordered systems
outside  the weak disorder limit could be described by the
the RMT model with non-integer symmetry parameter.
The same hyputhesis has been formulated in 
\cite{aniz}
on the basis of numerical  studies  of strongly anisotropic disordered systems 
This conclusion  also corresponds with  the hypothesis 
\cite{JPCM},
that any RMT which pretends to describe strongly localized regime,
should contain non-integer "symmetry parameter"
\be\label{beta}
\beta \sim\xi/L
\ee
where $L$ is the system size and $\xi$ is scaling parameter
(= localization length in the limit of strong disorder).

\medskip

In this Letter we  study the most simple random matrix model  
with  non-integer parameter $\beta$.
To characterize the statistics of transport, 
we  use quantities $z_i$
$i=1,2,\dots N$ ($N$ is the number of channels)  which
determine the eigenvalues $\lambda_i$ of the matrix $t^\dag t$ ($t$ is the transmission  matrix) as
$\lambda_i=(\cosh z_i -1)/2$. 
We calculate  the spectrum (the most probable values) of $z$ for $\beta$ being free parameter.
Then we apply  our solution to  $\beta$  given by (\ref{beta})  and show that 
obtained results correspond to the  spectrum of $z$'s for  
{\sl strongly disordered two dimensional} (2D) systems.

\medskip

We start with the RMT - form of the probability distribution of $z$s:
\be\label{one}
P(z)=\exp -\beta{\cal H}
\ee
\cite{pichard}
where ${\cal H}$ consists from one - particle potential 
and the two-particle interaction term.  In the limit of small $z$'s,  ${\cal H}$ has a
simple form
\be\label{interaction}
{\cal H} =\sum_i^N kz_i^2/2\beta-\log z_i/\beta-\sum_{i<j}\log|z_i^2-z_j^2|.
\ee
Following Muttalib
\cite{Muttalib},
the most probable values $\tilde{z}_i$ 
of $z$'s with distribution (\ref{one},\ref{interaction}), defined as the solution of the system
of nonlinear equations
\be\label{eq}
\frac{\partial{\cal H}}{\partial z_i}\Big|_{z_i=\tilde{z}_i}=0,
\ee
could be found analytically by method of Stieltjes 
through zeros of Laguerre polynomials $L_n^\alpha$
\cite{Mehta,Muttalib,JPCM} 
with
\be
\alpha = \frac{1}{\beta}-1.
\ee
Using properties of Laguerre polynomials
\cite{AS}, 
we express $\tilde{z}_i$ in the simple form
\cite{JPCM}
\be\label{spectrum}
\tilde{z}_i=\sqrt{\frac{1}{k(N+(\alpha+1)/2)}}~j_\alpha(i)\quad\quad N>>1
\ee
where $j_\alpha(i)$ is the $i$th zero of the Bessel function $J_\alpha$.

\medskip

In the special case $\beta=1$ parameter $\alpha=0$ and 
formula (\ref{spectrum})
describes correctly the spectrum of $z$'s in the metallic phase of orthogonal ensembles.
Indeed,   
$j_0(i)=(i\pi+1/2)/2$ for $i>>1$ so that $\tilde{z}_i\propto i$.  
It explains also
the small deviation from this linear behavior for small index $i$, 
($j_0(2)/j_0(1)=2.3$), which was observed numerically
\cite{JPCM}.

\medskip

Solution (\ref{spectrum}) has been derived without any restriction to value of $\beta$.
We can  apply  it also for $\beta\propto \xi/L<<1$. Then $\alpha>>1$.
In this limit we found zeros of Bessel functions in the form
\cite{AS}
\be\label{limit}
j_\alpha(i)\approx\alpha\left[1+\frac{2}{9}\left(\frac{3\pi}{8}\frac{4i-1}{\alpha}\right)\right]^{2/3}
\ee
Then, supposing that 
\be
k[N+(\alpha+1/2)]={\cal O}(L^0),
\ee
we obtain relation for the smallest $z_1$
\be\label{z1}
z_1= {\rm const}~\times L/\xi + {\cal O} (L^{1/3})
\ee
and for the difference $\Delta=z_2-z_1$
\be\label{diff}
z_2-z_1 \propto L^{1/3}.
\ee

\medskip

Formulae (\ref{z1},\ref{diff}) represent the main result of the present work.
To find their physical interpretation, we study 
the spectrum of $z$s for the strongly disordered two dimensional systems. 
Model is the standard 2D Anderson model with diagonal disorder. 
The strength of the disorder, $W=10$ and $=20$ assures that samples 
are in strongly localized regime.

Formula (\ref{z1}) gives correct $L$-dependence of $z_1$. It confirms  also
the physical interpretation of $\xi$ as the quantity proportional to the localization length.

Interpretation of the relation (\ref{diff}) is less trivial. However, our 
numerical data presented  in Figure 1, 
confirm that $z_2-z_1\propto L^\nu$ with exponent $\nu\approx 0.31$, 
in agreement with (\ref{diff}).

\smallskip

It is worth to note that the relation (\ref{diff}) does not hold in three dimensional
(3D) systems. In the last, we found that the difference $z_i-z_1$ is constant,
independent on the system size  and on the strength of the disorder.
This qualitative difference between 2D and 3D systems is caused by the dimension dependence
of the one particle potential, discussed in
\cite{JPCM}.

\medskip

In conclusion, we presented simple random matrix model which seems to describe
correctly transport properties of strongly disordered two-dimensional systems.
Our work has been inspired by previous success in the application of the random matrix theory to weakly disordered samples  and by recent observation
that the non-integer "symmetry parameter" $\beta$ could be, at least formally,
introduced in the theory
\cite{MutK}.  This observation requires the new physical interpretation
of $\beta$. We suggest that rather than being the symmetry parameter, $\beta$ represents
the "strength of the interaction" between particles described by "Hamiltonian"
(\ref{interaction}). Decrease of $\beta$ with increasing disorder just indicates
that the "interaction" is weaker. The symmetry of the model remains, of course,
unchanged.

We have not tried  to derive  the present random matrix model (without the $\cosh z$-term in the interaction) for the strongly disordered systems from the "first principles".
We only showed that simple quadratic model (\ref{one},\ref{interaction}) with $\beta$ given by (\ref{beta}) describes correctly some properties of the spectrum
of transfer matrix for two dimensional strongly disordered samples.
We found the present result inspiring for further progress in the generalization of the random matrix theory to system outside the weak disorder limit.

\bigskip

\noindent Acknowledgement: 
This work was  supported by Slovak Grand Agency, Grant VEGA n. 2/7174/20.\\

\newpage

\begin{figure}[t]
\centerline{ \psfig{file=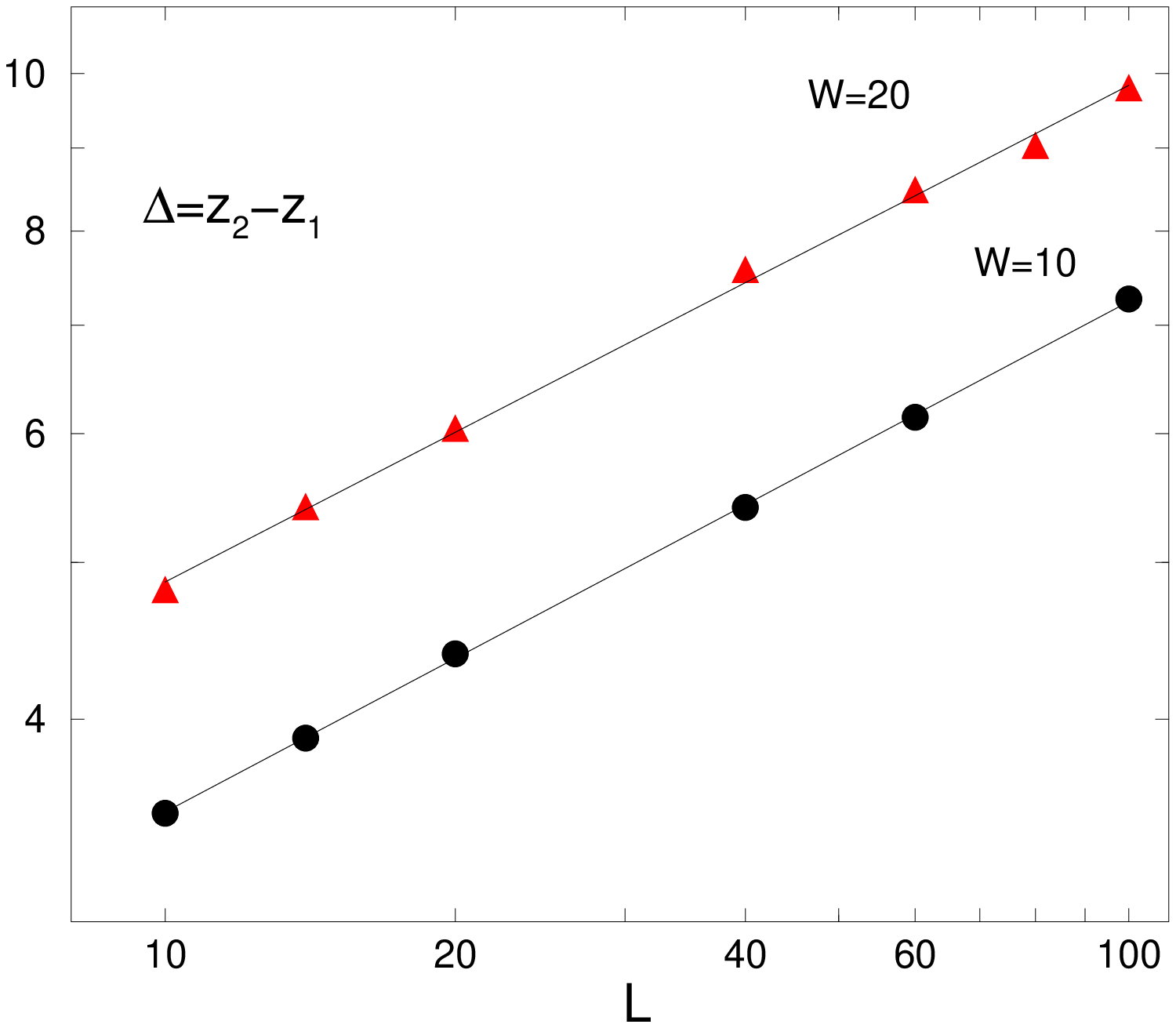,width=10cm}}
\caption{$L$-dependence of the difference $\Delta=z_2-z_1$ for
square samples with  disorder $W=10$
and $W=20$. Solid line is power fit $\Delta\sim L^\nu$ with $\nu\approx 0.31$.}
\end{figure}
\end{document}